\documentstyle[aps]{revtex}
\begin{document}
\voffset=-15mm
%
%
\title{MOMENT APPROACH FOR THE 2D ATTRACTIVE HUBBARD MODEL.}
%
\author{{\it J.J. Rodr\'{\i}guez-N\'u\~nez}, {\it C.E. Cordeiro} and 
{\it A. Delfino}}
\address{Instituto de F\'{\i}sica, Universidade Federal Fluminense,\\ 
Av.\ Litor\^anea S/N, 
Boa Viagem,\\ 24210-340 Niter\'oi - Rio de Janeiro, 
Brazil. \\e-m: jjrn@if.uff.br}

\date{\today}
\maketitle

%
%
\begin{abstract}
We constructed the one-particle spectral functions 
(diagonal and off-diagonal) which
reproduce BCS for weak coupling and which take 
into account the effect of
correlations on superconductivity in the attractive  
Hubbard model. The diagonal 
spectral function is composed of three 
peaks and the off-diagonal one is composed of two peaks. 
This ansatz satisfies the sum rules for the first six moments.  
Our solutions are valid for intermediate coupling, 
i.e., for $U/t \approx -4.0$. Our set of 
analytical equations for the unknown variables
is self-consistent and has been solved numerically in 
lowest order of the order parameter. As a result, we 
obtain that the presence of the third 
band, or {\it upper Hubbard band}, strongly renormalizes the 
two lower bands, making that the energy gap be  
${\bf k}$-dependent while the order parameter is pure 
s-wave. This shows that the order parameter and the 
gap are two different quantities.\\
\\
Pacs numbers: 74.20.-Fg, 74.10.-z, 74.60.-w, 74.72.-h
\end{abstract}

%
%
%
%

	The study of correlations has been renewed again by the
discovery of high $T_c$ superconducting oxide
materials (HTSC),\cite{Bednorz-Muller} since
these materials exhibit a short coherence
length, $\xi$, and a very large penetration depth, $\lambda$,
thus behaving like extreme type-II superconductors, i.e.
$\kappa \equiv {{\lambda}/{\xi}} \gg 1$.

    	We will use the on-site attractive Hubbard Hamiltonian
put forward by Micnas et al \cite{Micnas_et_al} as a
phenomenological model for describing the HTSC. Previous authors 
have used this model to study the bismuthate superconductors.\cite{3} 
The importance of the 
Hubbard model has been recognized by Denteneer et 
al\cite{Denteneer} who say  
that if the phase diagram of the Hubbard model were fully
understood, it might form the basis of understanding correlated
electrons as much as the Ising model did for understanding
critical phenomena. Schneider et al\cite{Schneider_et_al} have applied
this model to explain universal properties of several
families of these compounds, like the relation between transition
temperature, magnetic penetration depth and gap, at zero
temperature.  
It is likely that 
understanding this will provide insight into the 
effect of correlations on measurable quantities as 
shown by Singer et al\cite{Singer}. 

	We will use the exact relations of
Nolting\cite{Nolting} for the one-particle diagonal spectral
function together with the exact relations for the one-particle
off-diagonal spectral function, i.e., the anomalous Green's
function, to study systems with broken symmetry.\cite{KF}
We will start from the dynamical 
equations constructing one-particle Green's
functions which reduce to the BCS solution in the case of
weak coupling. Our solutions, even in the case of BCS, give
in a natural way the Hartree shift to the chemical potential. 

%
%

    	The Hamiltonian we study is the following
\begin{eqnarray}
\label{Ham}
H = - t\sum_{<i,j>\sigma}c_{i\sigma}^{\dagger}c_{j\sigma}
   + U \sum_i n_{i\uparrow}n_{i\downarrow}
   - \mu \sum_{i\sigma} n_{i \sigma}~~,
\end{eqnarray}
where $c_{i\sigma}^{\dagger}$($c_{i\sigma}$) are creation
(annihilation) electron operators with spin $\sigma$, $n_{i
\sigma} \equiv c_{i\sigma}^{\dagger}c_{i\sigma}$, $~t$ is a
hopping matrix element between the nearest sites $i$
and $j$, and $U$ is the onsite interaction. $\mu$ is the chemical
potential and we work in the grand canonical ensemble. In the
present study we consider an attractive interaction,
$U ~<~ 0$. The Hamiltonian
of Eq.~(\ref{Ham})~has been studied in
detail by Micnas et al\cite{Micnas_et_al}.

	Nolting\cite{Nolting} and Kalashnikov and
Fradkin\cite{KF} have derived exact relations for the one-particle
spectral functions, $A({\bf k},\omega)$ and 
$B({\bf k},\omega)$, which are given by:

\begin{eqnarray}\label{d0}
M^{(0)}({\bf k}) &\equiv&
\int_{-\infty}^{+\infty} A({\bf k},\omega) d\omega = 1~~~,\\
M^{(1)}({\bf k}) &\equiv& a_1 \equiv 
\int_{-\infty}^{+\infty} \omega A({\bf k},\omega) d\omega =
\varepsilon_{\bf k} - \mu +
\rho U ~~~, \\
M^{(2)}({\bf k}) &\equiv& a_2 \equiv 
\int_{-\infty}^{+\infty} \omega^2 A({\bf k},\omega) d\omega = 
(\varepsilon_{\bf k} - \mu)^2 +
2\rho (\varepsilon_{\bf k} - \mu) U + \rho U^2 ~~~,\\
M^{(3)}({\bf k})  &\equiv& a_3 \equiv 
\int_{-\infty}^{+\infty} \omega^3 A({\bf k},\omega) d\omega = \nonumber \\
(\varepsilon_{\bf k} - \mu)^3 &+&
3\rho (\varepsilon_{\bf k} - \mu)^2 U + 
(2 + \rho) \rho U^2 (\varepsilon_{\bf k} - \mu) +
\rho U^2 (1- \rho) (B_{\bar \sigma} - \mu) + \rho U^3 ~~~,\\
m^{(0)}({\bf k})  &\equiv&
\int_{-\infty}^{+\infty} B({\bf k},\omega) d\omega = 0~~~,\\
m^{(1)}({\bf k})  &\equiv&
\int_{-\infty}^{+\infty} \omega B({\bf k},\omega) d\omega = \Delta_o(T)~~~,
\end{eqnarray}

\noindent
where $\epsilon({\bf k}) =
-2 t (cos(k_xa) + cos(k_ya))$, with $a$ the lattice constant. 
$\Delta_o(T)$, 
the off-diagonal long range order, 
and $\rho$, the carrier concentration per spin, $n/2$, have 
to be calculated self-consistently. 
$A({\bf k},\omega)$
and $B({\bf k},\omega)$ are the diagonal and off-diagonal one-particle
spectral functions, respectively. They are defined by:

\begin{eqnarray}
A({\bf k},\omega) \equiv - \frac{1}{\pi} 
\lim_{\delta \rightarrow 0^+}Im[G({\bf k},\omega+i\delta)]~~; ~~~~~
B({\bf k},\omega) \equiv - \frac{1}{\pi} 
\lim_{\delta \rightarrow 0^+}Im[F({\bf k},\omega + i\delta)]~~. ~~~
\end{eqnarray}

	The parameter 
$B_{\bar \sigma}$, the Nolting's 
approximation,\cite{Nolting} is given in 
a self-consistent way.  
$\rho $ and $\Delta_o(T)$ are calculated 
from
\begin{equation}\label{density-spin}
\rho  = \frac{1}{N} \sum_{\bf k}\int_{-\infty}^{+\infty} d\omega
\frac{A({\bf k},\omega)}{e^{\beta \omega}+1}~~;~~~~~
\Delta_o(T) = \frac{1}{N} \sum_{\bf k}\int_{-\infty}^{+\infty} 
\frac{B({\bf k},\omega)}
{e^{\beta \omega}+1}~~~.
\end{equation}

    The main ingredient in any correct description of a physical
system is the availability of the one-particle 
Green's function. Of course,
in a many-body problem we cannot have a closed expression. 
In order to show the way, we first 
discuss the 2D BCS case in the framework of the moment approach 
and then we study the case of superconductivity beyond BCS, 
i.e., by including higher order moments. Here one must note that 
due to the Mermin-Wagner theorem no phase-transition is expected 
in 2D in systems with a continuous symmetry, and the formalism 
here is too simple to describe a Kosterlitz-Thouless 
phase-transition. Nonetheless, it is observed that the formalism 
does give a phase-transition.

\subsection{BCS case}

	Let us postulate as our 
one-particle spectral functions the following ones:

\begin{eqnarray}\label{BCS1}
A({\bf k},\omega) &=& \alpha_1({\bf k}) 
\delta (\omega - E_{\bf k}) +
\alpha_2({\bf k}) \delta (\omega + E_{\bf k})~~~,\\
B({\bf k},\omega) &=& \sqrt{\alpha_1({\bf k}) 
\alpha_2({\bf k})} \left[ \delta (\omega - E_{\bf k}) -
 \delta (\omega + E_{\bf k}) \right] ~~~,
\end{eqnarray}

	By using this Ansatz in the moment
equations, we obtain that:

\begin{eqnarray}\label{solutionsBCS}
E_{\bf k} = \sqrt{ (\varepsilon_{k} -\mu + \rho U )^2 + 
\Delta_o^2(T) }~~;~~~~
\alpha_1({\bf k}) = \frac{1}{2} \left(1 + 
\frac{\varepsilon_{\bf k} -\mu + \rho U}{E_{\bf k}} \right)~~;~~~~
\alpha_2({\bf k}) = 1 - \alpha_1({\bf k})~~~.
\end{eqnarray}

	Eqs.({\ref{solutionsBCS}) are clearly 
the BCS solutions with the Hartree shift,
$\rho U$,
included in a natural way. The quasi-particle excitation spectrum,
$E_{\bf k}$, has been obtained from 
the particular form of the off-diagonal single-particle
spectral function, 
i.e., the square root dependence. This
square root dependence will be kept 
when we include correlations, i.e., when
we include higher order moments, i.e., 
$a_2$ and $a_3$. This we will do next.

\subsection{Correlations on the BCS solution.}

	Now, we will go beyond 
the BCS solution by including correlations in
higher order. In this case, the physical
picture is that superconductivity 
is {\it similar} to BCS in the sense that we will have
quasi-particle spectra with a gap around the chemical potential.

	In order to keep the picture 
simple, we will assume that the chemical potential,
$\mu$, is close to the bottom of the 
free band. Then, the superconducting gap opens
up around the chemical potential 
and there is an upper Hubbard band which
remains almost similar to the Hubbard 
band when the order parameter is zero, i.e.,
when there is no symmetry breaking. 
In this case, we postulate that the
diagonal one-particle spectral density 
is composed of three poles as follows:

\begin{equation}\label{ansatz}
A({\bf k},\omega) = \alpha_1({\bf k}) 
\delta(\omega - \hat{\Omega}_1({\bf k})) +
\alpha_2({\bf k}) \delta(\omega - \hat{\Omega}_2({\bf k})) +
\alpha_3({\bf k}) \delta(\omega - \hat{\Omega}_3({\bf k}))~~~
\end{equation}

\noindent
In Eq. (\ref{ansatz}), the first two poles represent 
the behavior around the chemical potential and the
third pole is due to the influence 
of the upper Hubbard band, which in the 
case of the atomic limit, $t = 0$, is the upper Hubbard 
band or the 
band of single occupied states. The lower
band, now split in two, is the band of 
doubly occupied states for the case of almost 
atomic limit\cite{Micnas-et-al}. The
off-diagonal (anomalous) spectral function 
has the same form given as in the BCS case. 
Of course, our parameters $\alpha_j({\bf k})$ and
$\hat{\Omega}_j({\bf k})$  ($j = 1,2,3$), 
have to be calculated self-consistently ($\hat{\Omega}_1({\bf k}) 
= + E_{\bf k}$; $\hat{\Omega}_2({\bf k}) 
= - E_{\bf k}$; $\hat{\Omega}_3({\bf k}) 
= \Omega_{\bf k}$).  
Now, the parameter $B_{\bar \sigma}$ which 
appears in $a_3({\bf k})$ is generally
${\bf k}$-dependent. In the 
spherical approximation\cite{Nolting}, it is ${\bf k}$-independent and 
is calculated from the following equation:

\begin{equation}\label{narrowfactor}
\rho (1 - \rho) B_{\bar \sigma} = 
\frac{1}{N}\sum_{\bf k}\sum_{j=1}^{3}
\alpha_j({\bf k}) \varepsilon_{\bf k} 
\frac{1}{e^{\beta\hat{\Omega}_j({\bf k})}+1}
\left[ \frac{2}{U} (\hat{\Omega}_j({\bf k}) + \mu - 
\varepsilon_{\bf k}) - 1 \right] ~~~,
\end{equation}

\noindent
where $\beta = 1/(k_BT)$, $T$ the absolute 
temperature, $k_B$ the Boltzman's
constant, and $N$ the number of
lattice sites. Next, by putting our ansatz 
$A({\bf k},\omega)$ and $B({\bf k},\omega)$
into the set of moment equations, 
we get to lowest order in $\Delta^2_o$

\begin{eqnarray}\label{nalpha1-2}
\alpha_1({\bf k}) = \frac{1}{2} 
\left( \frac{\Omega_{\bf k} - 
a_1({\bf k})}{\Omega_{\bf k} - H_{\bf k}} \right) 
\left[ 1 + \frac{H_{\bf k}}{E_{\bf k}} \right]~;~~~
\alpha_2({\bf k}) = \frac{1}{2} \left( \frac{\Omega_{\bf k} - a_1({\bf k})}
{\Omega_{\bf k} - H_{\bf k}} \right) 
\left[ 1 - \frac{H_{\bf k}}{E_{\bf k}} \right]~;~~\alpha_3({\bf k}) 
= 1 - \alpha_1({\bf k}) - \alpha_2({\bf k})~~~,
\end{eqnarray}
\noindent
where the quasi-particle energy is given by
\begin{equation}\label{qp}
E_{\bf k} = \sqrt{ H_{\bf k}^2 + \left( \frac{\Omega_{\bf k} - 
H_{\bf k}}{\Omega_{\bf k} - a_1({\bf k})} \right)^2 
\Delta^2_o(T) }
\end{equation}
\noindent
with
\begin{eqnarray}\label{H_oOmega_o}
H_{\bf k} \approx H_o({\bf k}) &=& 
\Omega_1({\bf k}) = \frac{1}{2} \left[ \varepsilon_{\bf k} - 2\mu + U + B 
- \sqrt{\left( \varepsilon_{\bf k} - U - B \right)^2 + 
4\rho U (\varepsilon_{\bf k} - 
B) }\right] \nonumber \\
\Omega_{\bf k} \approx \Omega_o({\bf k}) &=& 
\Omega_2({\bf k}) = \frac{1}{2} \left[ \varepsilon_{\bf k} 
- 2\mu + U + B 
+ \sqrt{\left( \varepsilon_{\bf k} - U - B \right)^2 + 
4\rho U (\varepsilon_{\bf k} - 
B) }\right]~~~,
\end{eqnarray}

\noindent
where the $\Omega_i({\bf k})$'s are the ones which enter in the 
moment equations in the normal state (Eqs. 
(\ref{H_oOmega_o}) and Ref.\cite{Nolting}). However, due to the 
fact that the narrowing factor, $B$, is a sum over three 
frequencies of our problem (see Eq. (\ref{narrowfactor})) then 
all three parameters ($\mu$,$B$ and $\Delta(T)$) are mixed 
together. 

	From Eqs. (\ref{qp},\ref{H_oOmega_o}) we see that the 
quasi-particle energy, $E_{\bf k}$, has a gap which we find to be:
\begin{equation}\label{deltak}
\Delta(T,{\bf k}) = \Delta_o(T) \left( 
\frac{\Omega_2({\bf k}) - 
\Omega_1({\bf k})}{ \Omega_2({\bf k}) - a_1({\bf k})} \right) 
 ~~~
\end{equation}
	From Eq. (\ref{deltak}) we conclude that our gap equation, 
which is a manifestation of singularities in the density of states, 
is ${\bf k}$-dependent. Going back to our local Hamiltonian 
(see Eq. (\ref{Ham})), which in reciprocal space is a constant, then 
at the mean field level we should obtain a {\it pure s-wawe}, while 
going beyond mean field approximation we have been able to derive 
a $\bf k$-dependent gap. In our case, we have been able to 
modify the BCS results, from pure s-wave to 
a wave vector dependent energy gap.

	Our Ansatz is 
based on the assumption that the role of correlations is mainly 
taken into account in the diagonal one-particle spectral function. This 
implies that the off-diagonal order parameter 
has been taken to as the  
BCS one, i.e., to $m^{(1)}({\bf k})$. In a previous 
work,\cite{momentsbelow} it is found that both the diagonal and 
off-diagonal spectral functions have four peaks, symmetric in 
pairs, for 
$U/t = -4$. However, in the 
diagonal spectral function the fourth peak has a small weight and 
it is neglected here. 
If we want to 
consider two other peaks for the off-diagonal spectral function we 
have to include more moments, which is beyond the present work. 
Also, we have neglected life-time effects in Eq. (\ref{ansatz}), 
which would require the evaluation of higher moments. We leave this 
out from the present calculation.

In Fig. 1 we show the dependence of the reduced gap as function 
of reduced temperature. For 
comparison, we have also included 
the BCS gap. The critical temperature, $T_c/t$, has the  
value $0.80$ and the gap at zero temperature is  
$0.66t$. Then, for the ratio $\Delta(0)/T_c$ we find  
$0.825$. We should compare this with the BCS universal value, 
$1.76$,  for the same 
parameters\cite{9}-\cite{10}.   
In Fig. 2 we 
show the energy spectra for the three 
bands involved in the 
diagonal spectral density. We observe that 
the third band, $\Omega_{\bf k}$,  is 
almost always on top of the {\it 
superconducting bands}, i.e., 
$\pm E_{\bf k}$. In the inset of 
Fig. 2 we include a blowup of $E_{\bf k}$ close to 
its minimun, in order to calculate 
the gap. In Fig. 3 we present the ${\bf k}$ dependent 
factors, $\gamma_1({\bf k})$ and $\gamma_2({\bf k})$,
\begin{equation}\label{gamma12}
\gamma_1({\bf k}) = \frac{(\Omega_2({\bf k}) - \Omega_1({\bf k})) 
\Delta_o^2(T)}{(\Omega_2({\bf k}) - a_1({\bf k})) a_2({\bf k})}~~~; 
~~~\gamma_2({\bf k}) = \frac{(\Omega_2({\bf k}) - \Omega_1({\bf k}))
\Omega_1({\bf k})\Delta_o^2(T)}
{(\Omega_2({\bf k}) - a_1({\bf k}))a_3({\bf k})} 
 ~~~,
\end{equation}
for $T/t = 0.001, U/t = - 4.0$ and $\rho = 0.1$. These factors, 
which represent a correction 
to our frequencies, $H_{\bf k}$ and 
$\Omega_{\bf k}$, are zero at $T_c$. This implies that 
they do not change  
the value of $T_c$. However, for $T \approx 0$ they reach up to a  
.25. This contribution may look small 
but we have to remember that our equations are 
highly non-linear and a small change could produce 
a large change in the variables of 
interest, i.e., $\Delta(0)$.  
In consequence, our results support the calculations of 
Refs.\cite{9}-\cite{10} in the sense that the critical temperature 
is renormalized by more than a factor of two 
with respect to the BCS value. On the 
other hand, we argue that 
the value of the gap at zero temperature is going to 
be modified by the inclusion of the $\gamma (\bf k)$'s, 
which are going to increase the value of $\Delta(0)$. Due to 
these arguments, our calculation is of a perturbative character 
in $\Delta_o^2(T)$. In the calculation of Pedersen et al\cite{momentsbelow} 
the authors find a critical temperature of $T_c = 0.19$ by using a 
different moment approach where the spectral functions have 
four peaks. Our calculations for the spectral weight supports the  
view of ref\cite{momentsbelow} 
because there is an additional band which splits off from 
$\alpha_3({\bf k})$.\cite{elsewhere} 
The appearence of a fourth band will modify the 
value of $T_c$ and also the value of the gap. However, the 
numerical $Tc/t$ in the present work is different than the 
one given in ref.\cite{momentsbelow}. 

	By using the moment approach  
in the presence
of off-diagonal long range order, i.e.,
$\Delta_o(T) \neq 0$, to the negative 
Hubbard model, we have worked out the attractive 
Hubbard model in 2-D with an ansatz where the diagonal 
spectral function has three peaks and the off-diagonal 
one has two. The physical meaning of each of these 
peaks has been discussed. We have solved our 
equations in lowest order of $\Delta_o(T)$, an approximation 
which obviously is valid close to $T_c$ but it fails 
for zero temperature.  
We have seen 
that the effect of the third band is to renormalize the 
order parameter producing an energy gap which is 
{\bf k}-dependent. In consequence, in our approach the 
order parameter, $\Delta_o(T)$, and the energy gap, 
$\Delta(T,{\bf k})$, are not the same. A similar 
discussion was presented by Randeria, Duan and 
Shieh\cite{RDS}. The implications of this point is that we 
can shed  
light on the experiments carried out in the high-temperature  
superconductor materials since the experimentalists are 
talking about the symmetry of the order 
parameter and from the theoretical 
side we should refer to the energy gap. These two concepts 
are valid in mean field theory, i.e., in the BCS approximation. 
When we go beyond BCS as in the present work, it 
is difficult to asses the symmetry of the order parameter from 
tunelling experiments only.  
Then, 
from the experimental side it is difficult to find the 
symmetry of the order parameter 
unless the high temperature superconductors obey 
mean field equations.  
In short, the messages  
of this paper are 
\begin{itemize}
\item {\bf 1-} Due to correlations, the order parameter and the gap are 
two different quantities. In our case, from a local pairing attraction, 
i.e., pure s-wave in mean field, we get a $\bf k$ gap function.
\item {\bf 2-} Our calculations are of a perturbative character and they 
agree with Refs.\cite{9}-\cite{10}. Work is in progress\cite{elsewhere} 
to include additional bands in the diagonal spectral function. This 
will be published elsewhere. 
\end{itemize}

\section{Acknowlegments.}

    We would like to thank the Brasilian Agency CNPq 
(Project No. 300705/95-6) and CONICIT (project F-139) 
for finantial support. We profited from useful discussion with
Prof. R.R. dos Santos, Prof. B. Coqblin 
 and Dr. M.H. Pedersen. One of the 
authors (JJRN) wants to recognize that the result of this 
research is a product of a long discussion with the following  
collegues: Dr. M.H. Pedersen, Dr. 
J. Singer, Prof. H. Beck, Prof. T. Schneider and 
Prof. R. Micnas. 
We thank Mar\'{\i}a Dolores Garc\'{\i}a Gonz\'alez 
for reading the manuscript.
%
%
%
%

\vspace{1.2cm}

\begin{center}
{\huge Figures.}
\end{center} 
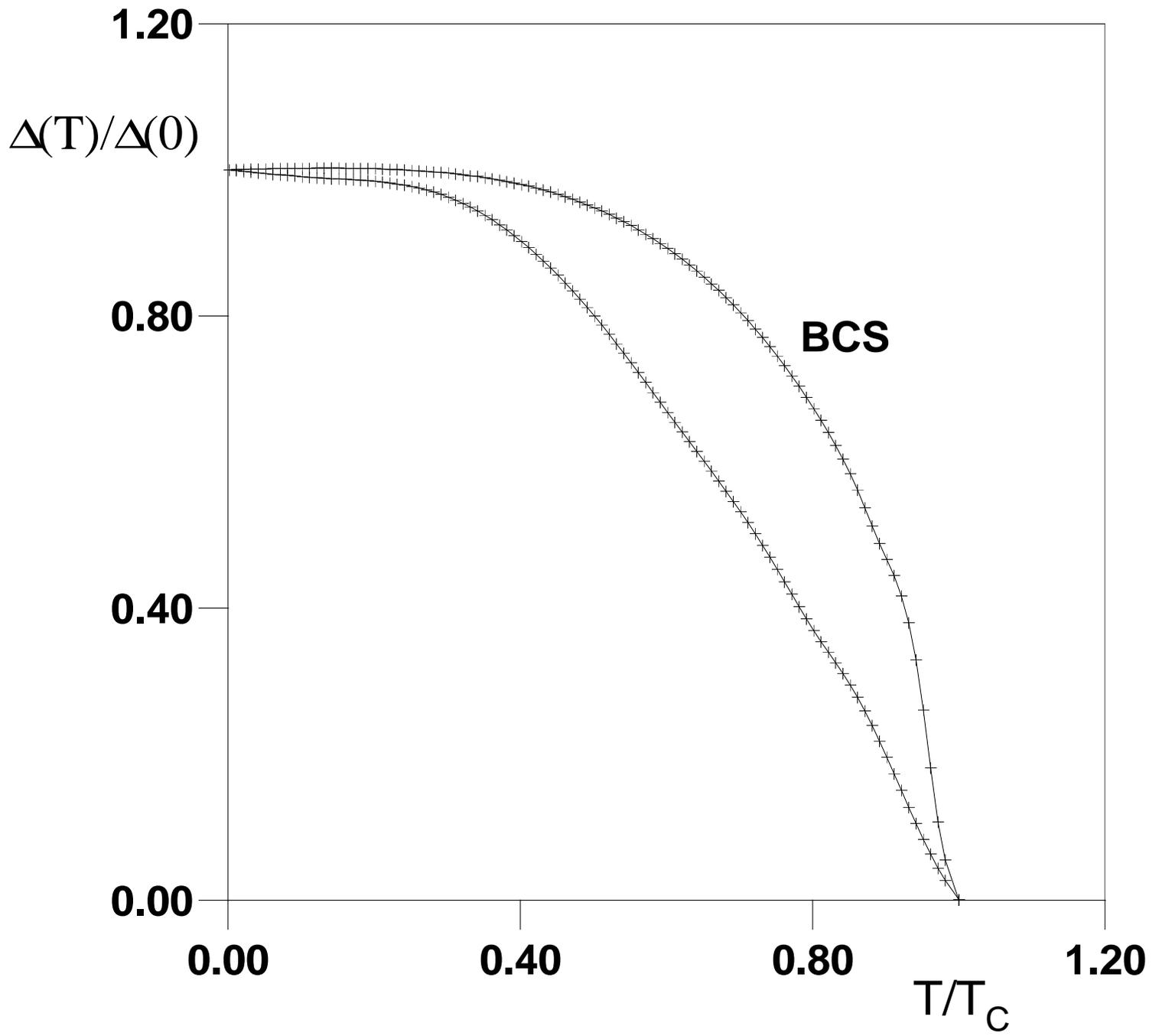
\noindent Fig. 1.- The energy reduced gap, 
$\Delta(T)/\Delta(0)$, as function of reduced temperature, 
$T/T_c$, for 
$U/t = -4.0$, $\rho = 0.1$. See Fig. 2 (inset) for the definition of 
$\Delta(T)$.\\

\noindent 
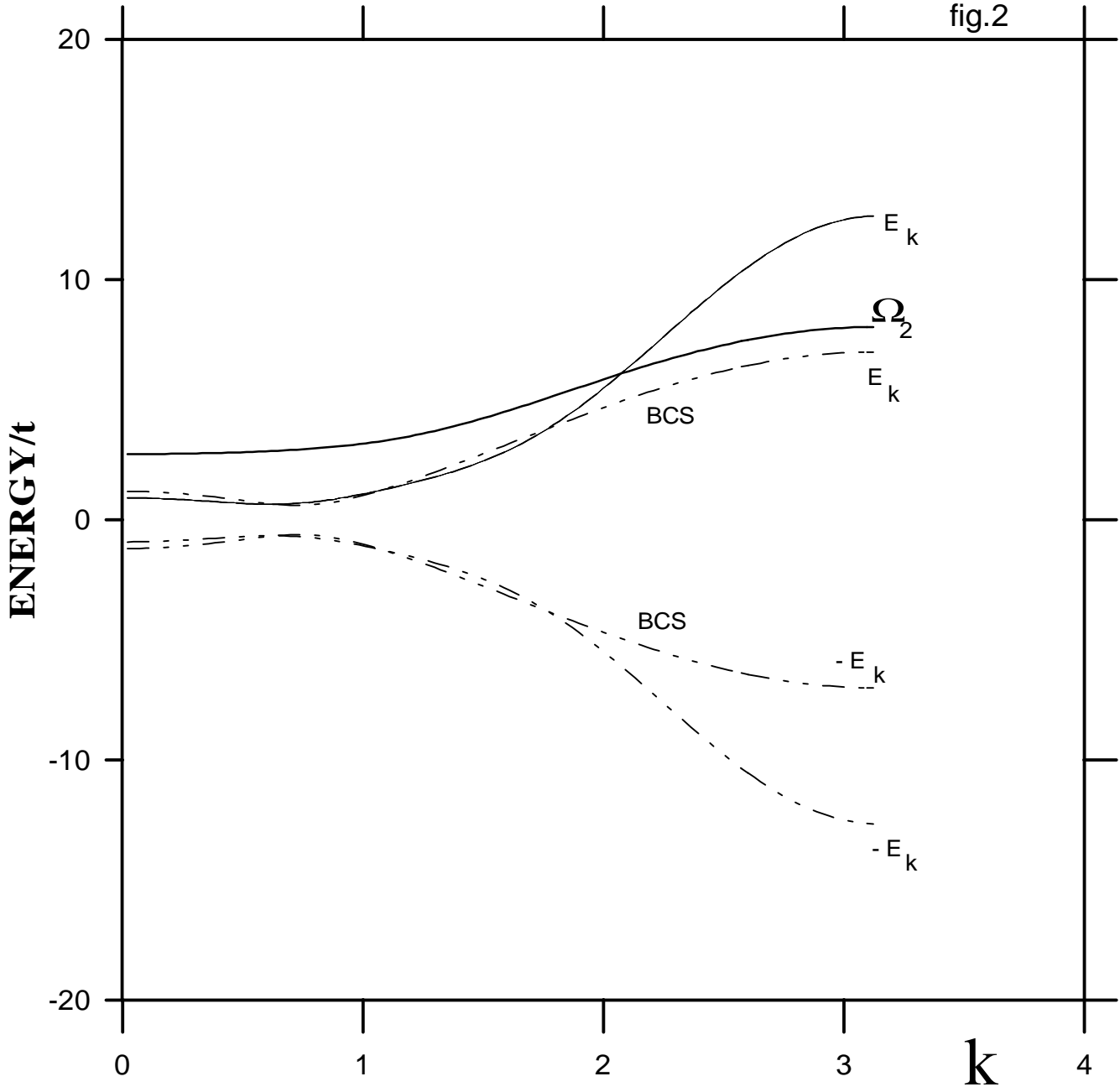
Fig. 2.- The energy bands, $\pm E_{\bf k}$ and $\Omega_{\bf k}$ 
along the diagonal of the Brillouin zone. For comparison, we have 
included the BCS energy bands. In the inset we have made a blow 
up of $+ E_{\bf k}$ close to the minimun of the band. This allows 
us to find $\Delta(T)$. Same parameters as in Fig. 1 with 
$T/t = 0.001$.\\

\noindent 
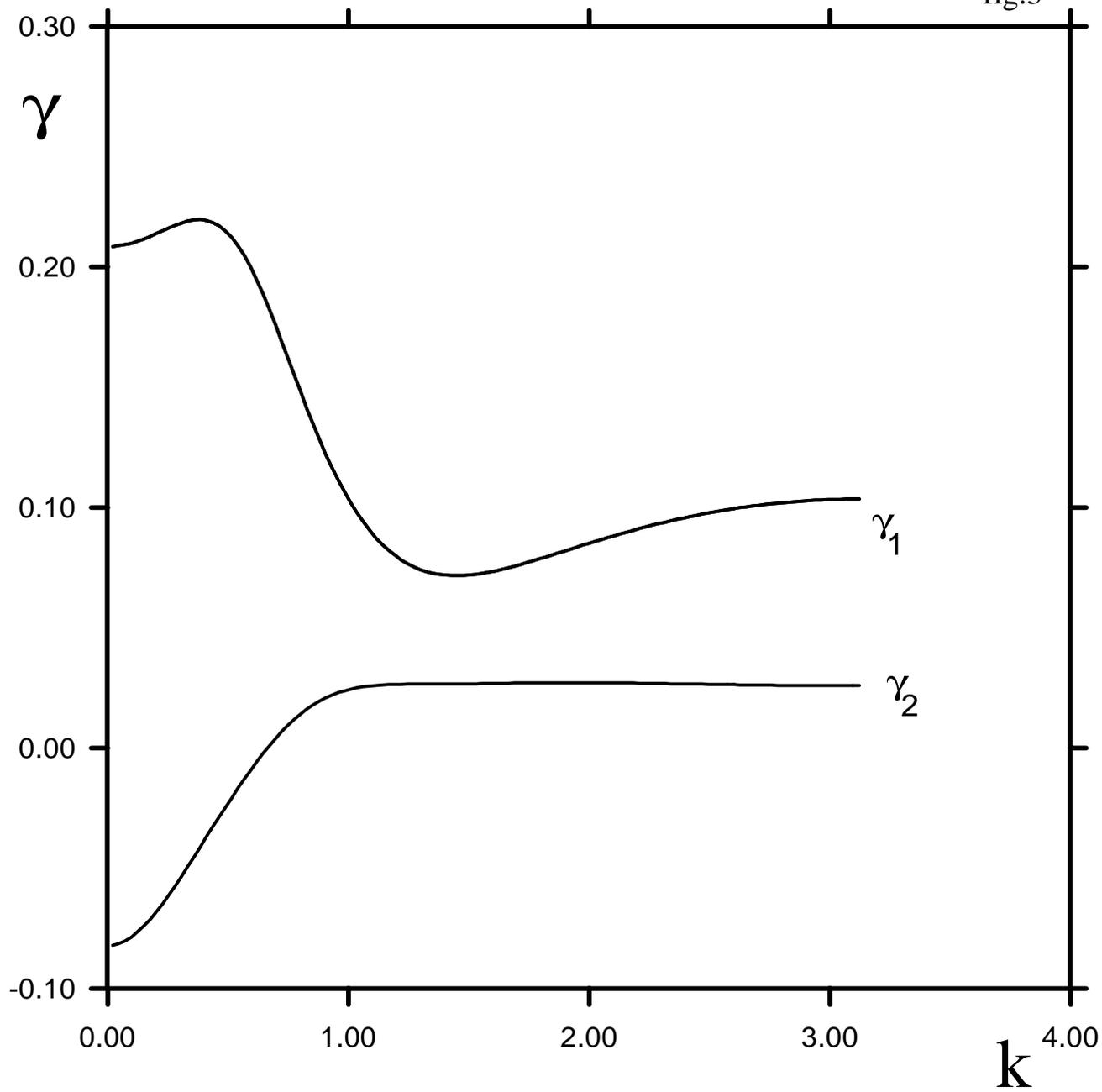
Fig. 3.- We plot $\gamma_1$,$\gamma_2$ vs k, along the diagonal 
of the Brillouin zone for the same parameters in Fig. 2. The 
parameters $\gamma$'s are defined in Eq. (19).\\

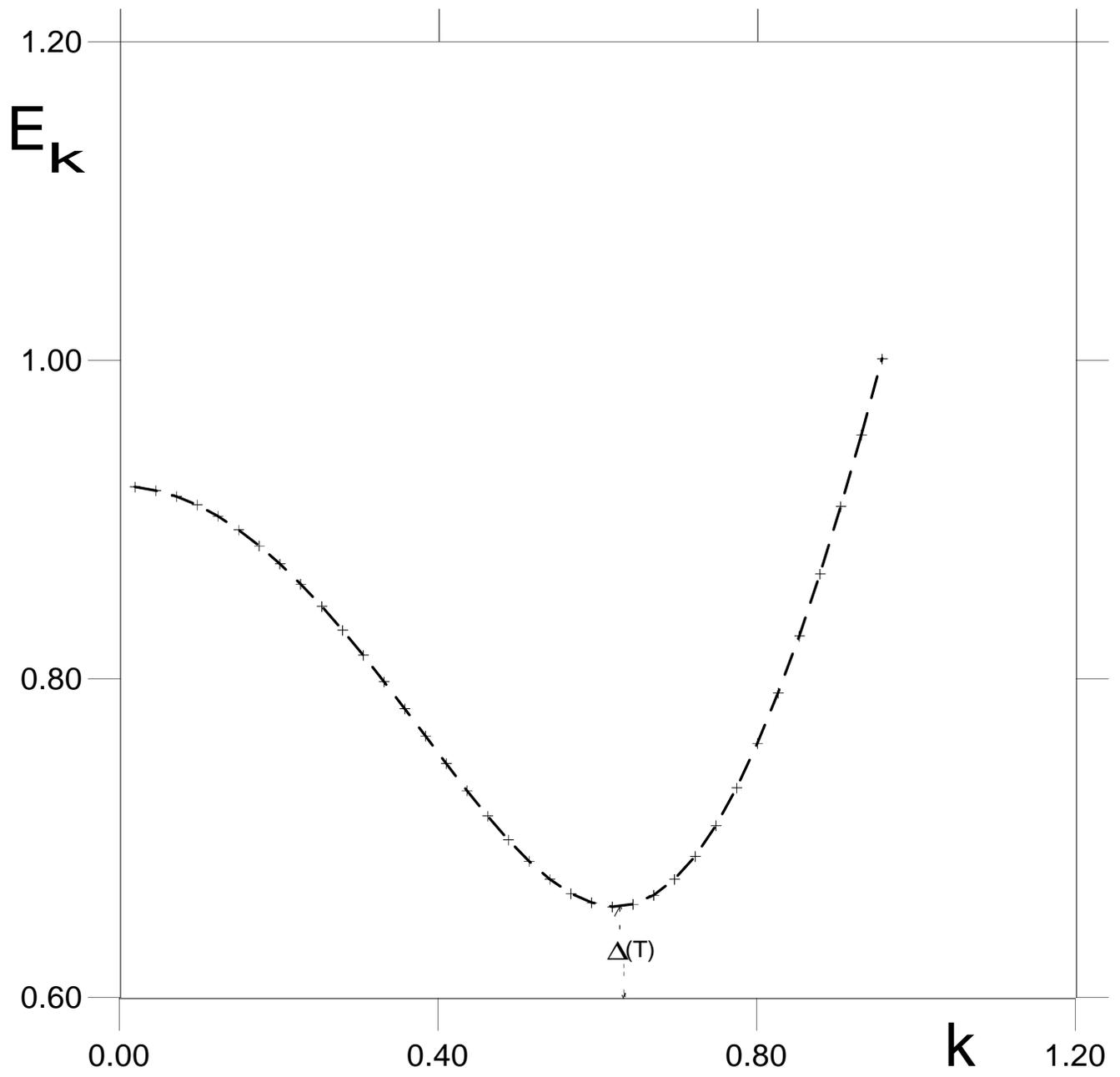
\end{document}